\documentclass[12pt,a4paper]{article}
\usepackage{epsfig}
\usepackage{amsmath}
\usepackage{latexsym}
\newtheorem{theorem}{Theorem}[section]
\newtheorem{proposition}[theorem]{Proposition}
\newtheorem{definition}[theorem]{Definition}
\newtheorem{lemma}[theorem]{Lemma}
\newtheorem{corollary}[theorem]{Corollary}
\newtheorem{example}[theorem]{Example}

\newcommand{\pf}{\noindent {\bf Proof: \ }}

\newcommand\abs[1]{\lvert#1\rvert}
\font\msbm=msbm10 at 12pt
\newcommand{\TT}{\mbox{\msbm T}}
\newcommand{\ZZ}{\mbox{\msbm Z}}
\newcommand{\FF}{\mbox{\msbm F}}
\newcommand{\RR}{\mbox{\msbm R}}
\font\msb=msbm10 at 8pt
\newcommand{\ZZZ}{\mbox{\msbm Z}}
\font\msb=msbm10 at 22pt
\newcommand{\Z}{\mbox{\msb Z}}
\begin{document}

\title{Bases of the Galois Ring $GR(p^r,m)$ \\ over the Integer Ring $\Z_{p^r}$}

\author{Virgilio Sison\thanks{This author gratefully acknowledges financial grant from the UPLB Diamond Jubilee-Development Fund Professorial Chair Award.}\\
Institute of Mathematical Sciences and Physics\\
University of the Philippines Los Ba\~{n}os\\
College, Laguna 4031, Philippines\\
Email: {\tt vpsison@uplb.edu.ph}}

\maketitle

\begin{abstract}
\boldmath The Galois ring $GR(p^r,m)$ of characteristic $p^r$ and cardinality $p^{rm}$, where $p$ is a prime and $r,m \ge 1$ are integers, is a Galois extension  of the residue class ring $\ZZZ_{p^r}$ by a root $\omega$ of  a monic basic irreducible polynomial of degree $m$ over $\ZZZ_{p^r}$. Every element of $GR(p^r,m)$ can be expressed uniquely as a polynomial in $\omega$ with coefficients in $\ZZZ_{p^r}$ and degree less than or equal to $m-1$, thus $GR(p^r,m)$ is a free module of rank $m$ over $\ZZZ_{p^r}$ with basis $\{1,\omega, \omega^2,\ldots, \omega^{m-1} \}$. The ring $\ZZZ_{p^r}$ satisfies the invariant dimension property, hence any other basis of $GR(p^r,m)$, if it exists, will have cardinality $m$. 

This paper was motivated by the code-theoretic problem of finding the homogeneous bound on the $p^r$-image of a linear block code over $GR(p^r,m)$ with respect to any basis. It would be interesting to consider the dual and normal bases of $GR(p^r,m)$.  

By using a Vandermonde matrix over $GR(p^r,m)$ in terms of the generalized Frobenius automorphism, a constructive proof that every basis of $GR(p^r,m)$ has a unique dual basis is given. The notion of normal bases was also generalized from the classic case for Galois fields. 

\end{abstract}

\section{Introduction}
\label{sect:int}
It was proved in \cite{abr} that every basis of the Galois ring $GR(4,m)$ has a dual basis, by treating each linear transformation from $GR(4,m)$ to $\ZZ_4$ as being uniquely determined in terms of the generalized trace function on $GR(4,m)$. In this paper this result is generalized to the Galois ring $GR(p^r,m)$ following the alternate method of MacWilliams and Sloane \cite{mac:slo} which constructs the dual basis using simple matrix algebra involving the generalized Frobenius automorphism. The material is organized as follows: Section~\ref{sect:pre} gives the preliminaries and basic definitions, while Section~\ref{sect:maj} gives the results.

\section{Preliminaries and definitions}
\label{sect:pre}
An overview of Galois fields and Galois rings, the Frobenius automorphism and the trace function, is presented in this section. For a thorough discussion of these topics, we refer the reader to \cite{mac}, \cite{mac:slo} and \cite{wan}. 

\subsection{Galois fields and Galois rings}

Let $p$ be a prime number and $r \ge 1$ an integer. Consider the ring $\ZZ_{p^r}$ of integers modulo $p^r$. When $r=1$ the ring $\ZZ_p$ with $p$ elements is a field and is usually denoted by $\FF_p$. Let  $\ZZ_{p^r}[x]$ be the ring of polynomials in the indeterminate $x$ with coefficients in  $\ZZ_{p^r}$. 

The Galois field with $p^m$ elements, denoted $\FF_{p^m}$, is a field extension $\FF_p[\alpha]$ of $\FF_p$ by a root $\alpha$ of an irreducible polynomial $\pi(x)$ of degree $m$ in $\FF_p[x]$.  Thus every element $z$ of $\FF_{p^m}$ can be expressed uniquely as a polynomial in $\alpha$ of the form  \begin{equation} \label{alp}
z = a_0 + a_1 \alpha + a_2 \alpha^2 + \ldots + a_{m-1} \alpha^{m-1} 
\end{equation}  with degree at most $m-1$ with the coefficients $a_i$ coming from $\FF_p$, and hence can also be written as an $m$-tuple $(a_0, a_1, \ldots, a_{m-1})$ in $\FF_p^m$.  Elements of $\FF_{p^m}$ may also be described as residue classes of the polynomials in $x$ with coefficients in $\FF_p$ reduced modulo $\pi(x)$. When $m =1$ we again have the field $\FF_p$.

The canonical projection homomorphism $\mu:\ZZ_{p^r} \rightarrow \FF_p$ is the mod-$p$ reduction map, and can be extended naturally as a map from $\ZZ_{p^r}[x]$ onto $\FF_p[x]$. This extended map is a  ring homomorphism with kernel $(p) = \ZZ_{p^r}[x]p = \{f(x)p \mid f(x) \in \ZZ_{p^r}[x] \}$.

Let $g(x)$ be a monic polynomial of degree $m \geq 1$ in $\ZZ_{p^r}[x]$. If $\mu(g(x))$ is irreducible in $\FF_p[x]$, then $g(x)$ is said to be {\it monic basic irreducible}. If $\mu(g(x))$ is primitive in $\FF_p[x]$, then $g(x)$ is said to be {\it monic basic primitive}. Clearly, monic basic primitive polynomials in  $\ZZ_{p^r}[x]$ are monic basic irreducible.

In the general sense, a {\it Galois ring} is a finite commutative local ring with identity $1 \ne 0$ such that the set of zero divisors together with the zero element forms the unique maximal principal ideal $(p1)$ for some prime number $p$.  The residue class ring $\ZZ_{p^r}[x]/(h(x))$, where $h(x)$ is a monic basic irreducible polynomial of degree $m$ in $\ZZ_{p^r}[x]$, is a Galois ring with characteristic $p^r$ and cardinality $p^{rm}$. The elements of $\ZZ_{p^r}[x]/(h(x))$ are residue classes of the form \begin{equation} a_0 + a_1x + \ldots + a_{m-1}x^{m-1} + (h(x)) \end{equation} where $a_i \in \ZZ_{p^r}$.  The identity is $1 + (h(x))$ and the zero element is $(h(x))$. The principal ideal $(p[1+(h(x))]) = (p + (h(x)))$ consists of all the zero divisors and the zero element, and is the only maximal ideal.

If $\deg h(x) = 1$ then $\ZZ_{p^r}[x]/(h(x))$ is the ring $\ZZ_{p^r}$. If $r =1$, the canonical homomorphism $\mu$ becomes the identity map and  $\ZZ_{p^r}[x]/(h(x)) = \FF_p[x]/(h(x)) \cong \FF_{p^m}$. Now let $\omega = x + (h(x))$, then $h(\omega)=0$ and every element $z$ of $\ZZ_{p^r}[x]/(h(x))$ can be expressed uniquely in the form 
\begin{equation} \label{add}
	z = a_0 + a_1 \omega + \ldots + a_{m-1} \omega^{m-1}
\end{equation}
where $a_i \in \ZZ_{p^r}$. We can thus think of $\ZZ_{p^r}[x]/(h(x))$ as a Galois extension $\ZZ_{p^r}[\omega]$ of $\ZZ_{p^r}$ by $\omega$. The elements take the {\it additive representation}  (\ref{add}), a generalization of (\ref{alp}) for $\FF_{p^m}$.  Since any two Galois rings of the same characteristic and the same cardinality are isomorphic, we simply use the notation $GR(p^r,m)$ for any Galois ring with characteristic $p^r$ and cardinality $p^{rm}$. 

The Galois ring $\mathcal R = GR(p^r,m)$ is a finite chain ring of length $r$, its ideals $p^i \mathcal R$  with $p^{(r-i)m}$ elements are linearly ordered by inclusion, \begin{equation} \{0\}=p^r \mathcal R \subset p^{r-1}\mathcal R \subset \ldots \subset p \mathcal R \subset \mathcal R \end{equation}
The quotient ring $\mathcal R / p\mathcal R \cong \FF_{p^m}$ is the residue field of $\mathcal R$. There exists a nonzero element $\xi$ of order $p^m-1$, which is a root of a unique monic basic primitive polynomial $h(x)$ of degree $m$ over $\ZZ_{p^r}$ and dividing $x^{p^m-1}-1$ in $\ZZ_{p^r}[x]$. Consider the set
\begin{equation} \label{tei}
\mathcal T = \{ 0,1,\xi,\xi^2, \ldots, \xi^{p^m-2} \}
\end{equation}
of T\"{e}ichmuller representatives. In this case, every element $z$ of $GR(p^r,m)$ has a unique {\it mutiplicative or $p$-adic representation} as follows
\begin{equation} \label{pad}
 z = z_0 + pz_1 + p^2z_2 + \ldots + p^{r-1}z_{r-1} 
\end{equation}
\noindent where $z_i \in \mathcal T$. We have that $z$ is a unit if and only if $z_0 \neq 0$, and $z$ is a zero divisor or 0 if and only if $z_0 = 0$. The units form a multiplicative group of order $(p^m-1)p^{(r-1)m}$, which is a direct product $\langle \omega \rangle \times \mathcal E$, where $\langle \omega \rangle$ is the cyclic group of order $p^m-1$ that is isomorphic to $\ZZ_{p^m-1}$ and $\mathcal E = \{1 + \pi \mid \pi \in (p)\}$ is a group of order $p^{(r-1)m}$. Let $\mu(\xi)=\alpha$.  It can be shown that $\alpha$ is a primitive element in $\FF_{p^m}$, and thus $\mu(\mathcal T)=\FF_{p^m}$. The $p$-adic representation in (\ref{pad}) is a generalization of the power representation of an element of $\FF_{p^m}$. 

We realize that $GR(p^r,m)$ is a free module of rank $m$ over $\ZZ_{p^r}$ with the set \begin{equation} \label{pol}
	\mathcal P_m(\omega) = \{ 1,\omega,\omega^2,\ldots,\omega^{m-1} \} 
	\end{equation} 
as a free basis. The set $\mathcal P_m(\omega)$ is called the {\it standard} or {\it polynomial basis} of $GR(p^r,m)$. The ring $\ZZ_{p^r}$  satisfies the invariant dimension property, hence any other basis of $GR(p^r,m)$, if it exists, will have cardinality $m$.  
\subsection{Generalized Frobenius automorphism and trace}

The {\it Generalized Frobenius map} $f$ on the Galois ring $\mathcal R=GR(p^r,m)$ is defined by  
\begin{equation} \label{fro}
 z^f := z_0^p + pz_1^p + p^2z_2^p + \ldots + p^{r-1}z_{r-1}^p 
\end{equation}
where $z$ has the $p$-adic representation given in (\ref{pad}). The map $f$ satisfies the following properties.
\begin{enumerate}
   	\item[(i)] $f$ is a ring automorphism of $\mathcal R$.
   	\item[(ii)]$f$ fixes every element of $\ZZ_{p^r}$. 
   	\item[(iii)]$f$ is of order $m$ and generates the cyclic Galois group of $\mathcal R$ over $\ZZ_{p^r}$.
\end{enumerate}

When $r=1$, the automorphism $f$ reduces to the usual Frobenius automorphism $\sigma$ of $\FF_{p^m}$ defined by $\sigma(z)=z^p$. 

The {\it generalized trace map} $T$ from $\mathcal R$ down to $\ZZ_{p^r}$ is given by 
\begin{equation} \label{gentra}
T(z) := z + z^f + z^{f^2} + \ldots + z^{f^{m-1}} \end{equation}
and satisfies the following properties.
\begin{enumerate}
	\item[(i)] $T$ is surjective and $\mathcal R/\ker T \cong \ZZ_{p^r}$.
	\item[(ii)] $T$ takes on each value of $\ZZ_{p^r}$ equally often $p^{r(m-1)}$ times.
   	\item[(iii)] $T(\alpha + \beta) = T(\alpha) + T(\beta)$ for all $\alpha,\beta \in \mathcal R$.
   	\item[(iv)]$T(\lambda\alpha) = \lambda T(\alpha)$ for all $\lambda \in \ZZ_{p^r}, \alpha \in \mathcal R$.
	\item[(v)]$T(\alpha^f) = (T(\alpha))^f = T(\alpha)$ for all $\alpha \in \mathcal R$.
\end{enumerate}

Again when $r=1$ the generalized trace map $T$ reduces to the trace map $t: \FF_{p^m} \rightarrow \FF_p$ defined by \begin{equation} \label{tra}
 t(\beta) = \beta + \beta^p + \beta^{p^2} + \ldots + \beta^{p^{m-1}} \: .
\end{equation}

\subsection{Homogeneous weight on $GR(p^r,m)$}

Let $R$ be a finite ring with identity $1 \ne 0$, and $\TT$ be the multiplicative group of unit complex numbers. The group $\TT$ is a one-dimensional torus. A {\it character} of $R$ (considered as an additive abelian group) is a group homomorphism $\chi: R \rightarrow \TT$. The set of all characters $\widehat{R}$ (called the {\it character module of $R$}) is a right (resp. left) $R$-module whose group operation is pointwise multiplication of characters and scalar multiplication is given by $\chi^r(x) = \chi(rx)$ (resp. $^r\!\chi(x) = \chi(xr)$). A character $\chi$ of $R$ is called a {\it right (resp. left) generating character} if the mapping $\phi: R \rightarrow \widehat{R}$ given by $\phi(r) = \chi^r$ (resp. $\phi(r) = \: ^r\!\chi$) is an isomorphism of right (resp. left) $R$-modules. The ring $R$ is called {\it Frobenius} if and only if $R$ admits a right or a left generating character, or alternatively, if and only if $\widehat{R} \cong R$ as right or left $R$-modules. It is known that for finite rings, a character $\chi$ on $R$ is a right generating character if and only if it is a left generating character. Further $\chi$ is a right generating character if and only if $\ker \chi$ contains no non-zero right ideals.

Let $\RR$ be the set of real numbers. We define a homogeneous weight on an arbitrary finite ring $R$ with identity in the sense of \cite{gre:sch:2}. Let $Rx$ denote the principal (left) ideal generated by $x \in R$. 

\begin{definition} \label{wei}
 A weight function $w:R \rightarrow \RR$ on a finite ring $R$ is called (left) homogeneous if $w(0) = 0$ and the following is true.
	\begin{itemize}
 	\item [(i)] If $Rx = Ry$, then $w(x) = w(y)$ for all $x,y \in R$.
	\item [(ii)] There exists a real number $\Gamma \ge 0$ such that 
		\begin{equation}
		\sum_{y \in Rx} w(y) = \Gamma \cdot \abs{Rx}, \:\:  \text{for all} \:  x \in R \setminus \{0\} \: . \label{ave} 
        	\end{equation}
	\end{itemize}
\end{definition}

Right homogeneous weights are defined accordingly. If a weight is both left homogeneous and right homogeneous, we call it simply as a homogeneous weight. The constant $\Gamma$ in (\ref{ave}) is called the {\it average value} of $w$. A homogeneous weight is said to be {\it normalized} if its average value is $1$. We can normalize the weight $w$ in Definition~\ref{wei} by replacing it with $\widetilde{w} = \Gamma^{-1}w$ \cite{hon:nec}. The weight $w$ is extended naturally to $R^n$, the free module of rank $n$ consisting of $n$-tuples of elements from $R$, via $w(z) = \sum_{i=0}^{n-1} w(z_i)$ for $z = (z_0,z_1,\ldots,z_{n-1}) \in R^n$. The homogeneous distance metric $\delta: R^n \times R^n \longrightarrow \RR$ is defined by $\delta(x,y)= w(x-y)$, for $x,y \in R^n$. 

It was proved in \cite{hon} that, if $R$ is Frobenius with generating character $\chi$, then every homogeneous weight $w$ on $R$ can be expressed in terms of $\chi$ as follows.
\begin{equation}\label{cha}
 w(x) = \Gamma \left[ 1 - \frac{1}{\abs{R^{\times}}} \sum_{u \in R^{\times}} \chi(xu)\right] 
\end{equation}
where $R^{\times}$ is the group of units of $R$. 

For the Galois ring $GR(p^r,m)$ we apply the following homogeneous weight given in \cite{gre:sch:1} for finite chain rings.  
\begin{gather}\label{whom}
   w_{\tt hom}(x) = \begin{cases}
		0 & {\rm if} \ x = 0 \\
		p^{m(r-1)} & {\rm if} \ x \in  \left(p^{r-1}\right) \setminus \{ 0 \} \\ 
		(p^m-1)p^{m(r-2)} & {\rm otherwise} 
	  \end{cases}
\end{gather}
where $\left(p^{r-1}\right)$ is the principal ideal generated by the element $p^{r-1}$ of $GR(p^r,m)$. Since the Galois ring $GR(p^r,m)$ is a commutative Frobenius ring with identity whose generating character is $\chi(z) = \xi^{b_{m-1}}$, where $\xi = \exp(2\pi i/p^r)$ for $z = \sum_{i=0}^{m-1} b_i \omega^i$, the weight (\ref{whom}) can be derived from (\ref{cha}). The group of units of $GR(p^r,m)$ has cardinality $p^{m(r-1)}(p^m-1)$ and it easy to compute from  (\ref{ave}) that its average value is equal to
\begin{equation}\label{gama} \Gamma = (p^m-1)p^{m(r-2)} \end{equation}
\noindent which is its minimum non-zero value. When $r=1$,  we have $\Gamma = (p^m-1)/p^m$  and $w_{\tt hom}$ is just the usual Hamming weight $w_{\tt Ham}$ on $\FF_{p^m}$. When $m=1$, the average value  is $\Gamma = (p-1)p^{r-2}$ for the integer ring $\ZZ_{p^r}$. 

\subsection{Codes over $GR(p^r,m)$}

A block code $C$ of length $n$ over an arbitrary finite ring $R$ is a nonempty subset of $R^n$. The code $C$ is called {\it right (resp. left) $R$-linear} if $C$ is a right (resp. left) $R$-submodule of $R^n$. If  $C$ is both left $R$-linear and right $R$-linear, we  simply call $C$ a linear block code over $R$. A $k \times n$ matrix  over $R$ is called a {\it generator matrix} of a linear block code $C$ if the rows span $C$ and no proper subset of the rows generates $C$.

Let the set $\mathcal B_m = \{ \beta_0, \beta_1, \ldots,\beta_{m-1}\}$ be a basis of the Galois ring $\mathcal R$ over $\ZZ_{p^r}$, and $B$ be a linear block code of length $n$ over $\mathcal R$.  We consider the map $\tau: \mathcal R\longrightarrow\ZZ_{p^r}^m$ defined by \begin{equation} \label{tau} \tau(z)=(a_0, a_1,\ldots, a_{m-1}) \end{equation} for $z = a_0 \beta_0 + a_1 \beta_1 + \ldots + a_{m-1}\beta_{m-1} \in \mathcal R$,  $a_i \in \ZZ_{p^r}$. This map is a bijection and can be extended coordinate-wise to $\mathcal R^n$. Thus, if $c \in B$ and $c = (c_0,c_1,\ldots,c_{n-1})$, $c_i = \sum_{j=0}^{m-1} a_{ij} \beta_j$, $a_{ij} \in \ZZ_{p^r}$, then \begin{equation}
\tau(c)=(a_{00},\ldots,a_{0,m-1},\ldots,a_{n-1,0},\ldots a_{n-1,m-1}) \end{equation}
in $\ZZ_{p^r}^{mn}$. The image $\tau(B)$ of $B$ under $\tau$ with respect to $\mathcal B_m$ is  called the {\it $p^r$-ary image} of $B$, and is obtained by simply substituting each element of $\mathcal R$ by the $m$-tuple of its coordinates over $B$. It is easy to prove that $\tau(B)$ is a linear block code of length $mn$ over $\ZZ_{p^r}$.  For the degenerate case $m=1$, the block code $B$ is a code over $\ZZ_{p^r}$ and the map $\tau$ is the identity map on $B$. We equip $\tau(B)$ with a homogeneous distance metric with respect to the weight $w_{\tt hom}$ as given in (\ref{whom}).

The following lemma from \cite{con:hei:hon} will be very useful in the succeeding discussion.

\begin{lemma} [Constantinescu, Heise and Honold, 1996] \label{lemma3} For any linear block code $C \subseteq \ZZ_{p^r}^n$ we have
    \begin{equation}
    \frac{w_{\tt hom}(C)}{\abs{C}} = \Gamma \cdot \abs{ \{i \mid \pi_i(C) \neq 0 \}}
    \nonumber
    \end{equation}
\noindent where $w_{\tt hom}(C)$ is the sum of the homogeneous weights of all codewords of %%@
$C$, and $\pi_i$ is the projection from $\ZZ_{p^r}^n$ onto the $i$-th coordinate.
\end{lemma}

\section{Major Results}
\label{sect:maj}
We denote by $w_{\tt hom}(S)$ the sum of the homogeneous weights of the elements of set $S$, that is, \begin{equation} \label{total} w_{\tt hom}(S) = \sum_{x \in S} w_{\tt hom}(x) \: .\end{equation} 

\begin{proposition} For any basis $\mathcal B_m = \{ \beta_0, \beta_1, \ldots,\beta_{m-1}\}$ of  $GR(p^r,m)$ over $\ZZ_{p^r}$ we have 
	\begin{equation} \label{anybasis}
		\sum_{x \in GR(p^r,m)} w_{\tt hom}(\tau(x)) = m(p-1)p^{rm+r-2} \: .
	\end{equation}
\end{proposition}

\pf Let $S = \{x \mid x \in GR(p^r,m) \}$. Then $\tau(S)$ is a linear block code over  $\ZZ_{p^r}$ of length $m$ and cardinality $p^{rm}$. Applying Lemma \ref{lemma3} to $\tau(S)$ gives us 
	\begin{equation}
		\frac{w_{\tt hom}(\tau(S))} {\abs{\tau(S)}} = \Gamma \cdot w_s(\tau(S)) \: .
		\nonumber
	\end{equation} 
	\noindent Therefore we have $w_{\tt hom}(\tau(S)) = \abs{\tau(S)} \cdot \Gamma \cdot w_s(S)$. The value of $\Gamma$ is given in (\ref{gama}), and the support size $w_s(\tau(S))$ of $\tau(S)$ is  $m$. Using the notation in (\ref{total}), the result now follows.

This proposition gives the simple corollary below which was used to prove the bound of Rabizzoni in \cite[Theorem 1]{rab}.   
\begin{corollary} For any basis $\mathcal B_m = \{ \beta_0, \beta_1, \ldots,\beta_{m-1}\}$ of $\FF_{p^m}$ over $\FF_p$ we have \[\sum_{x \in \FF_{p^m}} w_{\tt Ham}(\tau(x)) =  m(p-1)p^{m-1} \: . \]
\end{corollary}
\pf The Galois ring $GR(p,m)$ is the Galois field $\FF_{p^m}$, and the  homogeneous weight $w_{\tt hom}$ given in (\ref{whom}) is the Hamming weight $w_{\tt Ham}$ on $\FF_p$ with $\Gamma = (p-1)/p$.

The bound of Rabizzoni in \cite[Theorem 1]{rab} was extended to linear block codes over Galois rings in \cite{sol:sis}.  

Denote by ${\tt Mat}_m(\mathcal R)$ the ring of $m \times m$ matrices over the Galois ring $\mathcal R = GR(p^r,m)$. It is known that a matrix $A$ in ${\tt Mat}_m(\mathcal R)$ is nonsingular (or invertible) if and only if $\det A$ is a unit in $\mathcal R$. We will also use the usual notation $\abs{A}$ for the determinant of $A$. The matrix $A$ is {\it symmetric} if and only if $A = A^t$, and is {\it orthogonal} if and only if  $AA^t= A^tA=I$, where $A^t$ is the transpose of $A$ and $I$ is the identity matrix. We propose the following definition.

\begin{definition} \label{dua} Two bases $\{\alpha_i\} = \{\alpha_1,\alpha_2,\ldots,\alpha_m \}$ and $\{\beta_j\}=\{\beta_1,\beta_2,\ldots,\beta_m \}$ of $GR(p^r,m)$ over $\ZZ_{p^r}$ are said to be {\it dual}) if $T(\beta_i \alpha_j) = \delta_{ij}$, where $\delta_{ij}$ is the Kronecker delta.
\end{definition}
\begin{lemma} \label{van} The matrix $\Omega \in {\tt Mat}_m(\mathcal R)$ given by  \begin{eqnarray}
\Omega = \left(\begin{array}{ccccc} 1 & 1& 1 & \ldots & 1 \\ \omega & \omega^f & \omega^{f^2} & \ldots & \omega^{f^{m-1}} \\ \omega^2 & (\omega^2)^f & (\omega^2)^{f^2} & \ldots & (\omega^2)^{f^{m-1}} \\ \vdots & & & & \vdots \\ \omega^{m-1} & (\omega^{m-1})^f & (\omega^{m-1})^{f^2} & \ldots & (\omega^{m-1})^{f^{m-1}} \end{array}\right) \nonumber
\end{eqnarray}
is nonsingular.
\end{lemma}
\pf By the definition of the generalized Frobenius automorphism (\ref{fro}), it easy to show that $(\omega^j)^{f^i} = (\omega^{p^i})^j$ for $i,j=0,1,\ldots,m-1$. Hence, 
\begin{eqnarray}
\Omega = \left(\begin{array}{ccccc} 1 & 1& 1 & \ldots & 1 \\ \omega & \omega^p & \omega^{p^2} & \ldots & \omega^{p^{m-1}} \\ \omega^2 & {(\omega^p)}^2 & (\omega^{p^2})^2 & \ldots & (\omega^{p^{m-1}})^2 \\ \vdots & & & & \vdots \\ \omega^{m-1} & (\omega^p)^{m-1} & (\omega^{p^2})^{m-1} & \ldots & (\omega^{p^{m-1}})^{m-1} \end{array}\right)\nonumber
\end{eqnarray}   
which is a Vandermonde matrix whose determinant is  \begin{equation} \text {det } \Omega = \prod_{j=1}^{m-1} \prod_{i=j+1}^{m} (\omega^{p^{i-1}} - \omega^{p^{j-1}}) \end{equation} Each factor in this product is a unit of $\mathcal R$  so that $\det \Omega$ is a unit in $\mathcal R$.  

\begin{lemma} \label{aut} Let $\{\beta_j\} = \{\beta_1,\beta_2,\ldots,\beta_m \}$ be a basis. The matrix \begin{eqnarray} \label{autob}
B = \left(\begin{array}{ccccc} \beta_1 & \beta_1^f & \beta_1^{f^2} & \ldots & \beta_1^{f^{m-1}} \\  \beta_2 & \beta_2^f & \beta_2^{f^2} & \ldots & \beta_2^{f^{m-1}} \\  \vdots & & & & \vdots \\ \beta_m & \beta_m^f & \beta_m^{f^2} & \ldots & \beta_m^{f^{m-1}} \end{array}\right) \end{eqnarray} is invertible.
\end{lemma}
\pf Express the polynomial basis  $\mathcal P_m(\omega)$ in~(\ref{pol}) in terms of the basis $\{\beta_j\}$ as follows.  \begin{equation} \left( \begin{array}{c} 1 \\ \omega \\ \omega^2 \\ \vdots \\ \omega^{m-1} \end{array} \right)= \left( \begin{array}{cccc} a_{11} & a_{12} & \ldots & a_{1m} \\ a_{21} & a_{22} & \ldots & a_{2m} \\  a_{31} & a_{32} & \ldots & a_{3m} \\ \vdots & & & \vdots \\ a_{m1} & a_{m2} & \ldots & a_{mm}\end{array} \right)
  \left( \begin{array}{c} \beta_1 \\ \beta_2 \\ \beta_3 \\ \vdots \\ \beta_m \end{array} \right)
  \nonumber \end{equation} 
where $A = (a_{ij})$ is a nonsingular matrix over $\ZZ_{p^r}$. We compute the matrix product $AB$. The fact that the Frobenius automorphism $f$ fixes each $a_{ij}$ implies that $AB$ is the Vandermonde matrix $\Omega$. Hence by Lemma~\ref{van}, $\det AB$ is a unit in $\mathcal R$. Consequently, $\det B$ is a unit in $\mathcal R$.   

We shall call the matrix $B$ the {\it automorphism matrix} of $\mathcal R$ relative to the basis $\{\beta_j\}$.
\begin{corollary}
 $(\det B)^2$ is a unit in $\ZZ_{p^r}$.
\end{corollary}
\pf It can be shown that
\begin{eqnarray} \label{bbt}
BB^t = \left(\begin{array}{cccc} T(\beta_1^2) & T(\beta_1 \beta_2) & \ldots & T(\beta_1 \beta_m) \\
T(\beta_2 \beta_1) & T(\beta_2^2) & \ldots & T(\beta_2 \beta_m)  \\  \vdots & & & \vdots \\ T(\beta_m \lambda_1) & T(\beta_m \lambda_2) & \ldots & T(\beta_m^2) \end{array}\right)
\end{eqnarray}
which is a matrix over $\ZZ_{p^r}$. It follows that $(\det B)^2$ is an element of $\ZZ_{p^r}$. By Lemma~\ref{aut}, we get the result.

Of course, $\det B$ is not necessarily a unit in the base ring $\ZZ_{p^r}$, although it is a unit in $\mathcal R$ according to Lemma~\ref{aut}. Please see Example~\ref{exa:1}.
\begin{theorem} Every basis has a unique dual basis.
\end{theorem}
\pf We show the proof for $m=3$ without loss of essential generality. Let $\{ \beta_1, \beta_2,\beta_3\}$ be a basis, and consider the automorphism matrix \begin{eqnarray}
B = \left(\begin{array}{ccc} \beta_1 & \beta_1^f & \beta_1^{f^2}  \\ \beta_2 & \beta_2^f & \beta_2^{f^2} \\  \beta_3 & \beta_3^f & \beta_3^{f^2} \end{array}\right) \nonumber \end{eqnarray} which is nonsingular by Lemma~\ref{aut}. Let $\text{adj B} = (b_{ij})$ where $b_{ij} =(-1)^{i+j} \abs{B_{ji}}$. Then  \begin{eqnarray}
\text{adj B} = \left(\begin{array}{ccc} \lambda_1 & \lambda_2 & \lambda_3  \\ \lambda_1^f & \lambda_2^f & \lambda_3^f \\  \lambda_1^{f^2} & \lambda_2^{f^2} & \lambda_3^{f^2} \end{array}\right) \nonumber \end{eqnarray} 
where $\lambda_1\!=\!\beta_2^f\beta_3^{f^2}\!-\!\beta_2^{f^2}\beta_3^f$, $\lambda_2\!=\!\beta_1^{f^2} \beta_3^f\!-\!\beta_1^f\beta_3^{f^2}$, and $\lambda_3\!=\!\beta_1^f\beta_2^{f^2}\!-\!\beta_1^{f^2}\beta_2^f$
so that $B^{-1} = \abs{B}^{-1} \text { adj B}$. Note that \begin{eqnarray}
BB^{-1} = \abs{B}^{-1} \left(\begin{array}{ccc} T(\beta_1 \lambda_1) & T(\beta_1 \lambda_2) & T(\beta_1 \lambda_3) \\  T(\beta_2 \lambda_1) & T(\beta_2 \lambda_2) & T(\beta_2 \lambda_3) \\ T(\beta_3 \lambda_1) & T(\beta_3 \lambda_2) & T(\beta_3 \lambda_3) \end{array}\right)\nonumber
\end{eqnarray} Following the argument of \cite{abr} it can be shown by using the generalized trace  that $\{  \lambda_1/\abs{B}, \lambda_2/\abs{B}, \lambda_3/\abs{B} \}$ is a linearly independent set, and hence is the unique dual of $\{\beta_j\}$. 

\begin{example} \label{exa:1} The polynomial basis for $GR(4,2)$ is the set $\{1,\omega\}$ where $1 + \omega + \omega^2=0$. The automorphism matrix is  \begin{eqnarray}
B = \left(\begin{array}{cc} 1 & 1 \\  \omega & 3 + 3\omega \end{array}\right) \nonumber \end{eqnarray} with determinant $3+2\omega$ which is a unit in $GR(4,2)$. Observe that  $(3+2\omega)^2=1$ is a unit in $\ZZ_4$. The inverse \begin{eqnarray}
B^{-1} = \left(\begin{array}{cc} 3 + \omega & 1 + 2\omega \\  2 + 3\omega & 3 + 2\omega \end{array}\right) \nonumber \end{eqnarray} gives $\{3 + \omega, 1 + 2\omega\}$ as the dual of the polynomial basis.
\end{example}
\begin{example} The polynomial basis for $GR(4,3)$ is the set $\{1,\omega,\omega^2\}$ where $\omega$ is the root of the basic primitive polynomial $x^3+2x^2+x-1$ over $\ZZ_4$. The automorphism matrix is given by \begin{eqnarray}
B = \left(\begin{array}{ccc} 1 & 1 & 1 \\  \omega & \omega^2 & \omega^4 \\ \omega^2 & \omega^4 & \omega \end{array}\right) \nonumber \end{eqnarray} with determinant $3$. The inverse is given by  \begin{eqnarray}
B^{-1} =  \left(\begin{array}{ccc} \omega+3\omega^3 & w+3\omega^4 & w^2+3\omega^4 \\ \omega^2+3\omega^6 & 3\omega+\omega^2 & 3\omega+\omega^4 \\ \omega^4+3\omega^5 & 3\omega^2+\omega^4 & \omega+3\omega^2 \end{array}\right) \nonumber \end{eqnarray} so that $\{3+2\omega+2\omega^2, 2 + 2\omega + \omega^2, 2+\omega+2\omega^2\}$ is the dual basis. This corrects the mistake in \cite[Example 1]{abr}.
\end{example}
\begin{example} The dual of the polynomial basis for $\ZZ_8[\omega]$, where $\omega$ is the root of the basic primitive polynomial $7 + 5x + 6x^2+x^3$ over $\ZZ_8$, is the set $\{3+6\omega+6\omega^2$, $6+2\omega+5\omega^2$, $6+5\omega+2\omega^2\}$.
\end{example}

We apply Definition~\ref{dua} for the notion of self-dual basis. 
\begin{definition} The basis $\{\beta_1, \beta_2,\ldots, \beta_m\}$ is {\it self-dual} if $T(\beta_i\beta_j)=\delta_{ij}$. 
\end{definition}
\begin{definition} \label{normal} A normal basis of $GR(p^r,m)$ is a basis of the form $\{\alpha,\alpha^f,\alpha^{f^2},\ldots, \alpha^{f^{m-1}} \}$ where $\alpha \in GR(p^r,m)$ and $f$ is the generalized Frobenius automorphism given in (\ref{fro}). In this case we say that $\alpha$ generates $\beta_m$.  
\end{definition}

We have the following immediate results. 
\begin{theorem}
 Let $\{\beta_j\}$ be a basis with automorphism matrix $B$. Then $B$ is orthogonal if and only if $\{\beta_j\}$ is self-dual.
\end{theorem}
\pf From (\ref{bbt})  we get $BB^t = I$ $\Leftrightarrow$ $T(\beta_i \beta_j) = \delta_{ij}$. 
\begin{theorem}
 Let $\{\beta_j\}$ be a basis with automorphism matrix $B$. Then the following statements are equivalent.
 \begin{itemize}
 \item [(i)] The basis $\{\beta_j\}$ is a normal basis.
 \item [(ii)] The automorphism matrix $B$ is a symmetric matrix. 
 \item [(iii)] The Frobenius automorphism $f$ is the $m$-cycle $\beta_1 \mapsto \beta_2,\beta_2 \mapsto \beta_3,\ldots,\beta_m \mapsto \beta_1$.
 \end{itemize}
\end{theorem}
\pf This equivalence is evident from the construction of the automorphism matrix in~(\ref{autob}). The basis $\mathcal B_m = \{\beta_j\}$ is normal  $\Leftrightarrow$ $\beta_1$ generates $\mathcal B_m$, that is, $\beta_2 = \beta_1^f, \beta_3 = \beta_2^{f}= \beta_1^{f^2}, \beta_3 = \beta_2^{f} = \beta_1^{f^3},\ldots, \beta_{m-1}=\beta_{m-2}^f=\beta_1^{f^{m-1}}, \beta_m = \beta_{m-1}^f = \beta_1^{f^{m-1}}, \beta_m^f = \beta_1^{f^m} = \beta_1$ $\Leftrightarrow$ $B$ is symmetric $\Leftrightarrow$ $f$ is the $m$-cycle $\beta_1 \mapsto \beta_2,\beta_2 \mapsto \beta_3,\ldots,\beta_m \mapsto \beta_1$.  
\begin{example} The set $ \mathcal B_2 = \{\omega, \omega^2 = 3+3\omega \}$ is a normal basis for $\ZZ_4[x]/(x^2+x+1)$. The automorphism matrix relative to this basis is given by \begin{eqnarray} \left(\begin{array}{cc} \omega & 3+3\omega \\  3+3\omega & \omega \end{array}\right) \nonumber \end{eqnarray} which is not orthogonal, hence $\mathcal B_2$ is not self-dual. However $B$ is symmetric. \end{example}
\begin{example} The set $\mathcal B_3 = \{ 1+\omega, 1+\omega^2, 3+3\omega+3\omega^2\}$ of $GR(4,3) = \ZZ_4[x]/(x^3+2x^2+x+3)$ is a self-dual normal basis as the automorphism matrix is both orthogonal and symmetric. \end{example}

\end{document}